\documentclass{article} 
\usepackage{iclr2026_conference,times}
\usepackage{tikz}
\usepackage{pgfplots}
\pgfplotsset{compat=1.18}
\usepgfplotslibrary{groupplots}
\usepackage{changepage} 

\usepackage{amsmath,amsfonts,bm}

\def\eqref#1{equation~\ref{#1}}

\def\1{\bm{1}}

\DeclareMathAlphabet{\mathsfit}{\encodingdefault}{\sfdefault}{m}{sl}
\SetMathAlphabet{\mathsfit}{bold}{\encodingdefault}{\sfdefault}{bx}{n}

\usepackage{hyperref}
\usepackage{natbib}
\usepackage{url}
\usepackage{booktabs}
\usepackage{threeparttable}
\usepackage{makecell}
\usepackage{graphicx}
\usepackage{wrapfig}
\usepackage{amsmath,amsfonts,amssymb}
\usepackage{enumitem}

\title{RenoBench: A Citation Parsing Benchmark}

\author{parth sarin \\
Graduate School of Education \\
Stanford University\\
Stanford, California, USA \\
\texttt{psarin@stanford.edu}
\And
Juan Pablo Alperin \\
Public Knowledge Project \\
Simon Fraser University \\
Vancouver, British Columbia, Canada \\
\texttt{juan@alperin.ca}
\AND
Dione Mentis \\
DataCite \\
Hannover, Germany \\
\texttt{dione.mentis@datacite.org}
\And
Adam Buttrick \\
California Digital Library \\
University of California Office of the President \\
Oakland, California, USA \\
\texttt{Adam.Buttrick@ucop.edu}
}

\iclrfinalcopy 
\begin{document}

\maketitle

\begin{figure}[h]
    \centering
    \vspace{-2em}
    \includegraphics[width=0.7\linewidth]{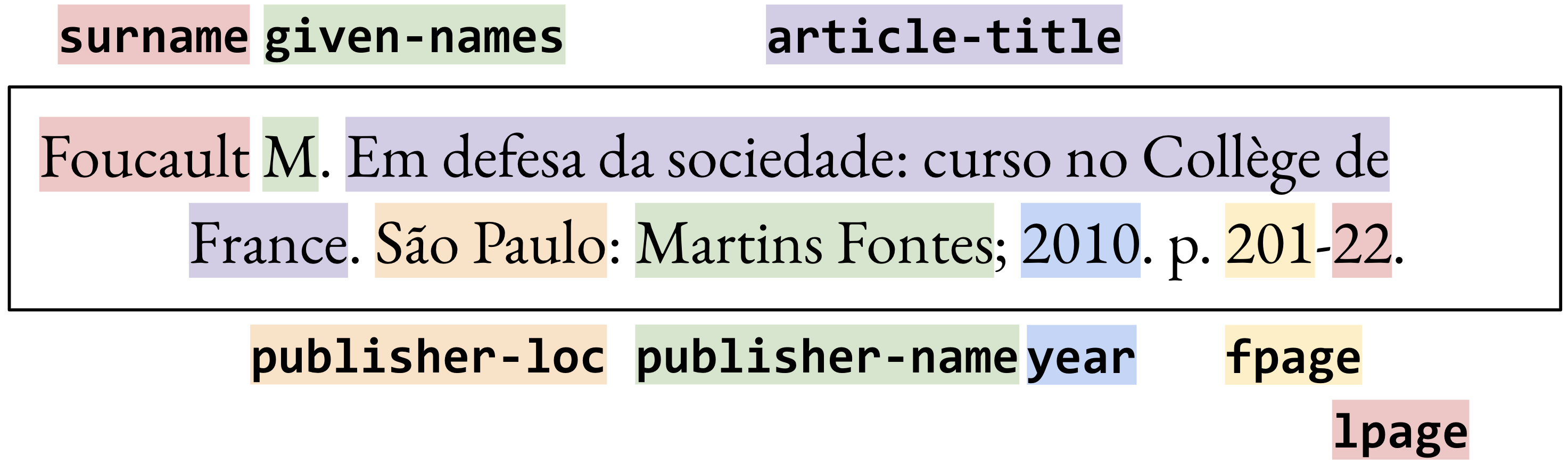}
    \vspace{-1em}
    \caption{An annotated citation from RenoBench}
    \label{fig:example-fraser-annotated}
    \vspace{-1em}
\end{figure}

\section{Introduction}

Despite sustained interest in machine readable citations \citep{OpenCitations,weimered,citex2026,wooc2025}, there is currently no community-accepted dataset to evaluate citation parsing. To assist the community working on this task, we introduce the \textbf{Re}ference An\textbf{no}tation Benchmark (RenoBench):\footnote{\url{https://huggingface.co/datasets/public-knowledge-project/ref-annotation-benchmark}} a standardized evaluation benchmark for citation parsing, which we define as the task of annotating a plain-text reference with structured components following the JATS standard \citep{NISO_JATS_Z39.96_2024}.

Existing benchmarks typically follow one of three patterns: First, many evaluations rely on small, hand-annotated datasets derived from a single discipline or venue (often computer science or biomedicine) such as those used to evaluate tools like GROBID and CERMINE \citep{lopez2009grobid,tkaczyk2018evaluation}. Second, recently, synthetic citation datasets like GIANT have emerged which are designed to maximize coverage of citation styles \citep{grennan2019giant}. These datasets have been shown to be useful in training contexts \citep{atanassova2020synthetic}, but synthetic references do not capture noise introduced by PDF extraction, multilingual syntax conventions, and publisher-specific production workflows encountered in practice \citep{lipinski2013evaluation}. Third, many citation parsers are evaluated on private datasets, limiting reproducibility \citep{tkaczyk2018evaluation}.

In contrast, RenoBench was assembled by extracting plain-text references from public domain PDFs from multiple journals. To our knowledge, RenoBench is the first public domain citation parsing benchmark sourced from real-world, multi-ecosystem conditions. With its release, we hope to complement existing datasets and enable reproducible comparison of citation parsing systems under realistic operating conditions \citep{atanassova2024breaking}.

\section{Methodology}
\subsection{Data Sources}
Article PDFs and their JATS XML annotations were derived from four publication platforms: SciELO, Redalyc, PKP, and Open Research Europe (ORE). The latter two, PKP and ORE, are publicly available \citep{garnett_xml_2016,openresearcheurope_corpus_2025}. JATS XML files from Redalyc and SciELO were obtained through communication with these platforms in February and April 2025, respectively.

We obtained the PDF files associated with the JATS XML for each corpus. ORE maintains a public index of XML and PDF URLs;\footnote{\url{https://open-research-europe.ec.europa.eu/published-xml-urls}, \url{https://open-research-europe.ec.europa.eu/about/developers}} the PKP corpus contains PDF versions of each article; and Redalyc and SciELO have publicly available PDFs at predictable URLs associated with each DOI. We then converted PDFs to markdown using \verb|markitdown| \citep{markitdown2025}, a lightweight conversion utility widely used to generate unstructured inputs to language models.

\subsection{Parsing and Matching}
Plain-text citations were extracted from the markdown using \verb|Llama-3.1-8B-Instruct|, programmatically verifying that the extracted citations appeared in the input text to prevent hallucinations. These plain-text citations were then matched to the annotated citations in the corresponding JATS XML files. Plain-text citations $p$ were matched to text-only (stripped) XML citations $c$ in order of decreasing similarity, where similarity is defined as \[ s(p, c) = 1 - \frac{d(p,c)}{\max{(|p|, |c|)}}, \] $d(p, c)$ is the edit distance between $p$ and $c$, and $|s|$ denotes the number of characters in a string $s$. All matches with similarity less than $0.75$ were excluded. This extraction and reconciliation procedure is described in more detail by \cite{sarin2025citation}. This matching resulted in a total of $161{,}625$ data points.

\subsection{Filtering}
We applied a suite of automated quality checks to remove citations with structural or content errors that would introduce noise into the benchmark. These checks fall into three categories:

\begin{enumerate}
    \item \textit{Structural validation checks} ensured well-formed XML (e.g., parseable markup, no namespace pollution) and sufficient annotation depth (citations with $\leq 2$ meaningful elements were excluded).
    \item \textit{Field-level validation checks} asserted regex-style heuristics against fields, detecting invalid dates (e.g., years with letter suffixes like ``2020a'', months $> 12$); problematic page numbers (e.g., reversed ranges where $\texttt{lpage} < \texttt{fpage}$, extreme values $> 10{,}000$ suggesting DOI fragments, or identical start and end pages); field confusion (e.g., years appearing in author fields, URLs in title fields); and author anomalies (e.g., initials stored as surnames, ``et al.'' parsed as an author name).
    \item \textit{Content consistency checks} aligned the structured annotation and the plain-text citation by fuzzy matching extracted surnames, sources, and article titles against the plain-text.
\end{enumerate}

Of the $161{,}625$ matched citation pairs, $115{,}193$ ($71.3\%$) passed all quality checks. The most common failure reasons were missing page numbers for journal articles ($27.7\%$ of errors) and reversed page number ranges ($14.4\%$). We excluded citation pairs that failed these checks.

\subsection{Sampling}
To construct a dataset capturing a range of contexts, we sought to balance the mix of cited languages, citing and cited publication types, presence of persistent identifiers (PIDs), and publication platform. To impute the cited article language, \verb|Llama-3.1-8B-Instruct| was used to return the two-letter ISO 639 language code based on the article title. The same model and input were also used to identify whether the citation contained a PID. Each citation was then represented as a binary feature vector encoding PID presence, preprint status, title language (one-hot), publication type (one-hot), and citation source (one-hot). 

Rather than sample uniformly, we optimized sampling weights to balance across these features using the following procedure: Let $N$ be the number of total citations and let $f_i \in \mathbb{R}^d$ be the feature vector for the $i$th data point, $x_i$. Take $\lambda \in \mathbb{R}^d$ to be a learnable collection of parameters. Assign a softmax sampling weight to each data point $w_i \gets \exp(\lambda^\intercal  f_i)$, so that the probability that $x_i$ is sampled in the first round is \[ p_i(\lambda) = \frac{\exp(\lambda^\intercal  f_i)}{\sum_{j=1}^N \exp(\lambda^\intercal  f_j)}.\] Then, define the aggregate feature vector to be $\mu(\lambda) = \mathbb{E}_{i \sim p} [f_i] = \sum_{i=1}^N p_i(\lambda) f_i$, intended to represent a ``soft'' sample, differentiable with respect to $\lambda$. A loss function $\mathcal{L}(\mu)$ was then designed to encourage adequate representation/balance. After $1{,}963$ steps at learning rate $10^{-3}$, convergence was reached and the final dataset, which contains $10,000$ citations, was sampled.

\subsection{Dataset Composition}
The resulting benchmark dataset has the following composition:
\begin{itemize}
  \item $59\%$ of citations include a persistent identifier (PID)
  \item $14\%$ of the citing articles are preprints
  \item Citations span eight languages, with English ($32\%$), Portuguese ($30\%$), and Spanish ($23\%$) most
common, followed by French ($7\%$), German ($3\%$), Italian ($2\%$), Russian ($2\%$), and Chinese ($1\%$)
  \item Publication types include journal articles ($53\%$), books ($30\%$), website pages ($8\%$), theses ($5\%$),
and conference proceedings ($4\%$)
  \item Documents are sourced from four platforms: SciELO ($47\%$), Redalyc ($24\%$), ORE ($14\%$), and PKP ($14\%$)
\end{itemize}

\section{Evaluation}
Several models were evaluated on RenoBench, including GROBID \citep{lopez2009grobid}, Qwen2.5 \citep{qwen_qwen25_2025}, Qwen3 \citep{yang2025qwen3technicalreport}, Mistral \citep{jiang_mistral7b_2023}, Gemma \citep{gemmateam2025gemma3technicalreport}, Llama \citep{grattafiori2024llama3herdmodels}, and GPT-OSS \citep{openai2025gptoss120bgptoss20bmodel}. The largest model tested was Qwen3-32B and the smallest was GROBID. We evaluated each of the language models by running inference on the Stanford Marlowe compute cluster \citep{kapfer_2025_14751899} with a prompt that provided two annotated examples of citations that were not included in the benchmark. GROBID was evaluated by querying its web API \href{https://kermitt2-grobid.hf.space/}{hosted on Hugging Face Spaces}. We also evaluated a fine-tuned version of \verb|Qwen3-0.6B| produced by \cite{cometadata2025citationparsing} following the procedure described in \cite{sarin2025citation}. Table~\ref{tab:citation-recall} shows the recall accuracy of different models for different markup fields.

We believe that precision is an unreliable metric on this dataset because the JATS XML often omits valid markup elements that models correctly produce. We evaluated all of the models for precision and manually reviewed 330 randomly sampled errors. We found that correct annotations were being marked as incorrect due to incomplete ground truth 92\% of the time for the \textit{year} field, 85\% for the \textit{source} field, and 72\% for the \textit{article-title} field. For completeness, we still report precision in Table~\ref{tab:citation-precision}.

Tables~\ref{tab:citation-recall} and~\ref{tab:citation-precision} report accuracy where, even if the model produced malformed XML, we still considered assertions for fields if they appeared between valid XML tags (e.g., \verb|<year>2026</year>| may be asserted even if the \verb|<mixed-citation>| tag is not closed). Tables~\ref{tab:citation-recall-strict} and~\ref{tab:citation-precision-strict} report stricter recall and precision metrics, which do not consider those assertions, such that field-level accuracy is bounded above by the valid XML percentage.

Finally, we attempted to optimize the prompt using GEPA \citep{agrawal2025gepareflectivepromptevolution}, using recall on a held-out training subset as the optimization objective. Prompt optimization did not yield consistent or meaningful improvements over the baseline prompts for any of the evaluated models so we do not report these results.

\begin{table}[htbp]
\begin{adjustwidth}{-1.5cm}{-1.5cm} 
\centering
\caption{Recall Across Models and Fields}
\label{tab:citation-recall}
\vspace{0.2em}
\begin{threeparttable}
\small
\setlength{\tabcolsep}{5pt} 
\begin{tabular}{l|r|rrrr|rrr}
\toprule
\textbf{Model} & \textbf{Valid XML} & \textbf{Surname} & \textbf{Given} & \textbf{Title} & \textbf{Source} & \textbf{Volume} & \textbf{Issue} & \textbf{Year} \\
\midrule
GROBID & \textbf{100.0} & 71.6 & 18.8 & 65.2 & 51.8 & 90.0 & 86.8 & \textbf{97.9} \\
Qwen/Qwen2.5-3B-Instruct & 83.5 & 93.0 & 64.1 & 93.9 & 54.3 & 95.5 & 83.8 & 93.8 \\
Qwen/Qwen2.5-7B-Instruct & 87.6 & \textbf{95.8} & 74.5 & \textbf{94.0} & 53.3 & 96.0 & 86.9 & 90.6 \\
Qwen/Qwen3-0.6B & 81.4 & 87.3 & 62.7 & 84.7 & 50.6 & 92.4 & 92.7 & 91.7 \\
\makecell[l]{Qwen/Qwen3-0.6B\\\,+ cometadata/citation-parsing-lora} & 95.0 & 92.6 & \textbf{77.9} & 93.4 & \textbf{73.3} & \textbf{98.6} & 95.3 & 97.2 \\
Qwen/Qwen3-8B & 78.2 & 86.6 & 70.9 & 82.0 & 51.2 & 90.6 & 89.8 & 80.7 \\
Qwen/Qwen3-32B & 78.4 & 86.8 & 69.2 & 84.7 & 55.6 & 92.3 & 90.5 & 86.2 \\
google/gemma-3-1b-it & 27.7 & 84.9 & 52.5 & 44.5 & 37.4 & 70.9 & 64.0 & 69.8 \\
meta-llama/Llama-3.1-8B-Instruct & 80.3 & 82.4 & 58.6 & 92.7 & 57.3 & 98.1 & 96.0 & 94.5 \\
mistralai/Mistral-7B-Instruct-v0.2 & 86.0 & 85.4 & 69.0 & 91.9 & 52.8 & 95.8 & 94.3 & 91.6 \\
openai/gpt-oss-20b & 88.3 & 89.1 & 67.0 & 93.1 & 57.4 & 98.1 & \textbf{96.5} & 96.7 \\
\bottomrule
\end{tabular}

\begin{tablenotes}
\footnotesize
\item All values represent recall (\%) on field extraction from bibliographic citations.
\item \textbf{Bold} indicates best performance in each column.
\item Valid XML shows the percentage of examples producing well-formed, parseable XML.
\item Field extraction uses a parser with regex fallback to recover fields from malformed XML.
\item Edit distance thresholds: surname/year/volume/issue = 0 (exact), source = 5, title = 10.
\end{tablenotes}
\end{threeparttable}
\end{adjustwidth}
\end{table}

\section{Conclusion}
RenoBench provides the first public domain benchmark for evaluating citation parsing under real-world scholarly publishing conditions. Our hope is that it will enable more meaningful and transparent comparisons of reference markup systems that are grounded in references that are noisy, heterogeneous, and multilingual. In presenting our RenoBench, our own evaluations highlight how language models perform on this task relative to established tools and demonstrate the promise of small, domain-specific language models for citation parsing. We see RenoBench as a foundation for future work on improved models and downstream citation indexing applications. If the benchmark proves useful, we propose that it be expanded with higher-fidelity ground truth (i.e., with more detailed JATS tagging) to support more reliable precision evaluation.

\newpage
\bibliography{iclr2026_conference}
\bibliographystyle{iclr2026_conference}

\newpage
\appendix
\section{Appendix}
\begin{table}[htbp]
\begin{adjustwidth}{-1.5cm}{-1.5cm} 
\centering
\caption{Precision Across Models and Fields}
\label{tab:citation-precision}
\vspace{0.2em}
\begin{threeparttable}
\small
\setlength{\tabcolsep}{5pt} 
\begin{tabular}{l|r|rrrr|rrr}
\toprule
\textbf{Model} & \textbf{Valid XML} & \textbf{Surname} & \textbf{Given} & \textbf{Title} & \textbf{Source} & \textbf{Volume} & \textbf{Issue} & \textbf{Year} \\
\midrule
GROBID & \textbf{100.0} & 70.2 & 18.7 & 58.4 & 48.0 & 69.8 & 49.1 & 96.9 \\
Qwen/Qwen2.5-3B-Instruct & 83.5 & 86.5 & 62.5 & 56.0 & 53.4 & 86.7 & \textbf{88.0} & 97.7 \\
Qwen/Qwen2.5-7B-Instruct & 87.6 & 89.0 & 72.3 & 59.2 & 53.0 & 87.1 & 78.3 & 98.2 \\
Qwen/Qwen3-0.6B & 81.4 & 79.5 & 59.0 & 51.1 & 47.4 & 42.9 & 25.7 & 95.1 \\
\makecell[l]{Qwen/Qwen3-0.6B\\\,+ cometadata/citation-parsing-lora} & 95.0 & 83.6 & 73.5 & 55.7 & \textbf{66.9} & 43.5 & 25.2 & 96.4 \\
Qwen/Qwen3-8B & 78.2 & \textbf{91.8} & \textbf{76.1} & 60.2 & 59.1 & 82.3 & 74.5 & 91.4 \\
Qwen/Qwen3-32B & 78.4 & 89.5 & 71.8 & 60.9 & 59.7 & 86.4 & 78.2 & 92.4 \\
google/gemma-3-1b-it & 27.7 & 81.3 & 49.8 & 41.1 & 48.2 & 57.6 & 34.1 & 94.5 \\
meta-llama/Llama-3.1-8B-Instruct & 80.3 & 79.2 & 58.3 & 60.7 & 59.1 & 85.4 & 74.3 & \textbf{98.4} \\
mistralai/Mistral-7B-Instruct-v0.2 & 86.0 & 85.2 & 69.3 & \textbf{61.0} & 65.2 & \textbf{88.7} & 71.7 & 97.8 \\
openai/gpt-oss-20b & 88.3 & 85.1 & 65.8 & 60.1 & 57.7 & 88.4 & 84.8 & 98.2 \\
\bottomrule
\end{tabular}

\begin{tablenotes}
\footnotesize
\item All values represent precision (\%) on field extraction from bibliographic citations.
\item \textbf{Bold} indicates best performance in each column.
\item Valid XML shows the percentage of examples producing well-formed, parseable XML.
\item Precision measures the fraction of generated fields that are correct (avoiding hallucinations).
\item Edit distance thresholds: surname/year/volume/issue = 0 (exact), source = 5, title = 10.
\end{tablenotes}
\end{threeparttable}
\end{adjustwidth}
\end{table}

\begin{table}[htbp]
\begin{adjustwidth}{-1.5cm}{-1.5cm} 
\centering
\caption{Recall (Strict) Across Models and Fields}
\label{tab:citation-recall-strict}
\vspace{0.2em}
\begin{threeparttable}
\small
\setlength{\tabcolsep}{5pt} 
\begin{tabular}{l|r|rrrr|rrr}
\toprule
\textbf{Model} & \textbf{Valid XML} & \textbf{Surname} & \textbf{Given} & \textbf{Title} & \textbf{Source} & \textbf{Volume} & \textbf{Issue} & \textbf{Year} \\
\midrule
GROBID & \textbf{100.0} & 71.6 & 18.8 & 65.2 & 51.8 & 90.0 & 86.8 & \textbf{97.9} \\
Qwen/Qwen2.5-3B-Instruct & 83.5 & 79.0 & 54.9 & 81.6 & 47.0 & 84.1 & 73.5 & 79.2 \\
Qwen/Qwen2.5-7B-Instruct & 87.6 & 83.7 & 65.2 & 83.4 & 47.1 & 85.0 & 76.7 & 79.8 \\
Qwen/Qwen3-0.6B & 81.4 & 72.5 & 52.9 & 72.7 & 44.2 & 80.8 & 81.8 & 75.8 \\
\makecell[l]{Qwen/Qwen3-0.6B\\\,+ cometadata/citation-parsing-lora} & 95.0 & \textbf{88.5} & \textbf{74.6} & \textbf{89.3} & \textbf{70.1} & \textbf{94.3} & \textbf{91.0} & 92.8 \\
Qwen/Qwen3-8B & 78.2 & 72.5 & 60.3 & 73.0 & 45.6 & 78.0 & 77.6 & 70.5 \\
Qwen/Qwen3-32B & 78.4 & 74.0 & 59.5 & 75.9 & 50.0 & 81.0 & 78.0 & 74.2 \\
google/gemma-3-1b-it & 27.7 & 21.8 & 13.5 & 7.6 & 6.1 & 11.9 & 11.0 & 13.8 \\
meta-llama/Llama-3.1-8B-Instruct & 80.3 & 65.7 & 45.6 & 69.6 & 43.8 & 73.7 & 72.5 & 76.0 \\
mistralai/Mistral-7B-Instruct-v0.2 & 86.0 & 74.6 & 60.6 & 80.9 & 46.8 & 85.3 & 82.5 & 79.4 \\
openai/gpt-oss-20b & 88.3 & 78.7 & 59.1 & 80.8 & 49.2 & 83.0 & 82.1 & 85.9 \\
\bottomrule
\end{tabular}

\begin{tablenotes}
\footnotesize
\item All values represent recall (\%) on field extraction from bibliographic citations.
\item \textbf{Bold} indicates best performance in each column.
\item Valid XML shows the percentage of examples producing well-formed, parseable XML.
\item The strict parser rejects malformed XML (mismatched tags, unclosed elements), returning empty results.
\item Edit distance thresholds: surname/year/volume/issue = 0 (exact), source = 5, title = 10.
\end{tablenotes}
\end{threeparttable}
\end{adjustwidth}
\end{table}

\begin{table}[htbp]
\begin{adjustwidth}{-1.5cm}{-1.5cm} 
\centering
\caption{Precision (Strict) Across Models and Fields}
\label{tab:citation-precision-strict}
\vspace{0.2em}
\begin{threeparttable}
\small
\setlength{\tabcolsep}{5pt} 
\begin{tabular}{l|r|rrrr|rrr}
\toprule
\textbf{Model} & \textbf{Valid XML} & \textbf{Surname} & \textbf{Given} & \textbf{Title} & \textbf{Source} & \textbf{Volume} & \textbf{Issue} & \textbf{Year} \\
\midrule
GROBID & \textbf{100.0} & 70.2 & 18.7 & \textbf{58.4} & 48.0 & 69.8 & 49.1 & \textbf{96.9} \\
Qwen/Qwen2.5-3B-Instruct & 83.5 & 73.4 & 53.5 & 48.7 & 46.2 & 76.4 & \textbf{77.1} & 82.5 \\
Qwen/Qwen2.5-7B-Instruct & 87.6 & 77.8 & 63.2 & 52.5 & 46.8 & 77.1 & 69.1 & 86.5 \\
Qwen/Qwen3-0.6B & 81.4 & 66.1 & 49.8 & 43.9 & 41.4 & 37.5 & 22.6 & 78.6 \\
\makecell[l]{Qwen/Qwen3-0.6B\\\,+ cometadata/citation-parsing-lora} & 95.0 & \textbf{79.9} & \textbf{70.4} & 53.3 & \textbf{64.1} & 41.6 & 24.0 & 92.0 \\
Qwen/Qwen3-8B & 78.2 & 76.8 & 64.7 & 53.6 & 52.6 & 70.8 & 64.4 & 79.8 \\
Qwen/Qwen3-32B & 78.4 & 76.3 & 61.7 & 54.6 & 53.7 & 75.8 & 67.4 & 79.6 \\
google/gemma-3-1b-it & 27.7 & 20.8 & 12.8 & 7.0 & 7.9 & 9.6 & 5.8 & 18.6 \\
meta-llama/Llama-3.1-8B-Instruct & 80.3 & 63.1 & 45.4 & 45.6 & 45.2 & 64.2 & 56.2 & 79.1 \\
mistralai/Mistral-7B-Instruct-v0.2 & 86.0 & 74.5 & 60.9 & 53.7 & 57.7 & \textbf{79.0} & 62.7 & 84.8 \\
openai/gpt-oss-20b & 88.3 & 75.1 & 58.0 & 52.2 & 49.4 & 74.8 & 72.2 & 87.2 \\
\bottomrule
\end{tabular}

\begin{tablenotes}
\footnotesize
\item All values represent precision (\%) on field extraction from bibliographic citations.
\item \textbf{Bold} indicates best performance in each column.
\item Valid XML shows the percentage of examples producing well-formed, parseable XML.
\item The strict parser rejects malformed XML (mismatched tags, unclosed elements), returning empty results.
\item Edit distance thresholds: surname/year/volume/issue = 0 (exact), source = 5, title = 10.
\end{tablenotes}
\end{threeparttable}
\end{adjustwidth}
\end{table}
\end{document}